\documentclass[nofootinbib,amsmath,aps,prl,reprint]{revtex4-2}
\usepackage{amssymb} 
\usepackage{graphicx}
\usepackage{times}
\usepackage{slashed}
\usepackage{bm}
\usepackage{color}

\begin{document}
\title{Alternative interpretation of relativistic time-reversal and the time arrow}
\author{T. Zalialiutdinov$^{1,\,2}$, D. Solovyev$^1$, D. Chubukov$^{2,\,3}$, S.  Chekhovskoi$^{1}$  and L. Labzowsky $^{1,\,2}$}

\affiliation{ 
$^1$ Department of Physics, St. Petersburg State University, Petrodvorets, Oulianovskaya 1, 198504, St. Petersburg, Russia\\
$^2$ Petersburg Nuclear Physics Institute named by B.P. Konstantinov of National Research Centre "Kurchatov Institut", St. Petersburg, Gatchina 188300, Russia\\
$^3$ Saint Petersburg State Electrotechnical University, Professora Popova Street, 197376, St. Petersburg, Russia
}

\begin{abstract}
It is well-known that the 4-rotation in the 4-dimensional space-time is equivalent to the CPT-transformation (C is the charge conjugation, P is the space inversion and T is the time-reversal). The standard definition of the T-reversal includes the change of the sign of time variable and replacement of the initial state of the particle (system of particles) by the final state and vice versa. Since the time-reversal operation changes the state of a particle, the 
particle's wave function cannot be the eigenfunction of the corresponding operator with a certain eigenvalue, as in the case of space parity. Unlike the CPT-transformation, the separate P, T, or C transformations cannot be reduced to any 4-rotation. The extended Lorentz group incorporates all the separate C, P, or T transformations which do not bring the time axis out of the corresponding light cone. The latter restriction is included in the standard definition of the time-reversal. In the present communication, we ignore this restriction. 

This allows to introduce the "time arrow" operator and characterize every particle by the new quantum number - "time arrow" value. The wave functions of all particles are eigenfunctions of this operator with eigenvalues equal to "time arrow" values. The particles with the "time arrow" values opposite to the "time arrow" value in our universe form another universe (anti-universe). The existence of anti-universe can be confirmed, in principle, by laboratory (atomic) experiments. The anti-universe may be also considered as a candidate to the role of dark matter.
\end{abstract}

\maketitle

The time invariance (invariance under the time-reversal) of all the basic equations of motion in classical and quantum physics remains one of the most long-standing problems in fundamental physics. The indirect evidence of T-violation based on the CPT-theorem (C is the charge conjugation, P is the space inversion and T is the time-reversal operations) is the observation of CP-violation in the decay of heavy mesons \cite{1}. The search for the universal T-violating interactions in nature started in 1950 in the paper \cite{str40} where it was suggested to observe the electric dipole moment (EDM) of the neutron in the magnetic resonance experiment. The existence of EDM of any particle which is not truly neutral
(and neutron is not truly neutral) violates both P- and T-invariance. Later it was found theoretically that the electron EDM (eEDM) should be greatly enhanced in heavy atoms and especially in heavy diatomic molecules \cite{2,3,4,5}. In \cite{3,5} it was also shown that apart from the interaction of eEDM with an external electric field another P, T-odd interaction should exist in atomic systems: a scalar-pseudoscalar P, T-odd interaction between atomic electron and the nucleus. In an external electric field both P, T-odd interactions lead to the same linear Stark shift of atomic levels and can not be distinguished in any experiment on a certain species. The most advanced recent experiments on the observation of P, T-odd interactions
in heavy diatomic molecules, \cite{6} are still very far above the values predicted by the Standard Model (SM) \cite{7,8,9}. The first direct observation of T-noninvariance \cite{1a} is performed again in meson physics. 

It is important to understand what happens to the particles which undergo such interaction, e.g. with the electron in an atom. Trying to understand this, we have found that the answer is not possible within the standard interpretation of T-reversal in relativistic physics and needs the alternative interpretation which is the subject of the present communication. This alternative interpretation provides a new contribution to the recent studies on the time symmetry in theoretical cosmology and results in the possible existence of the anti-universe as a candidate for the dark-matter what can be in principle confirmed by laboratory (atomic) experiments. 

It is well-known that the CPT transformation in 4-dimensional space-time can be reduced to the 4-rotation \cite{10}. However, separate C, P, T transformations can not be reduced to 4-rotation. The standard extended Lorentz group incorporates all these separate C, P, T, transformations. However, the transformations which bring the time axis out of the upper (lower) light cone are forbidden \cite{10} (see Fig. 1 (a)). An alternative interpretation of time reversal in relativistic theory consists of allowing such transformations (see Fig. 1 (b)). The corresponding Lorentz transformation we will call additionally extended Lorentz transformation  \footnote{In \cite{weinbergbook} the Lorentz group which includes the additionally extended Lorentz transformations is called full Lorentz group}. We should stress, that all the details of the standard extended Lorentz group theory remain untouched if we consider any transformations allowed in the standard theory. Note that the conclusion of the observation of T-noninvariance in \cite{1a} was made on the basis of the standard interpretation of time reversal. However, the allowance of the additional transformations in Fig. 1 (b) leads to the arrival of new properties, absent in the standard theory.

\begin{figure}[hbtp]
\caption{The 2-dimensional images of the 4-dimensional light cones. (a) the time axis transformations allowed in the standard extended Lorentz group. (b) Additional transformations of the time axis in the additionally extended Lorentz group.}
\centering
\includegraphics[scale=0.8]{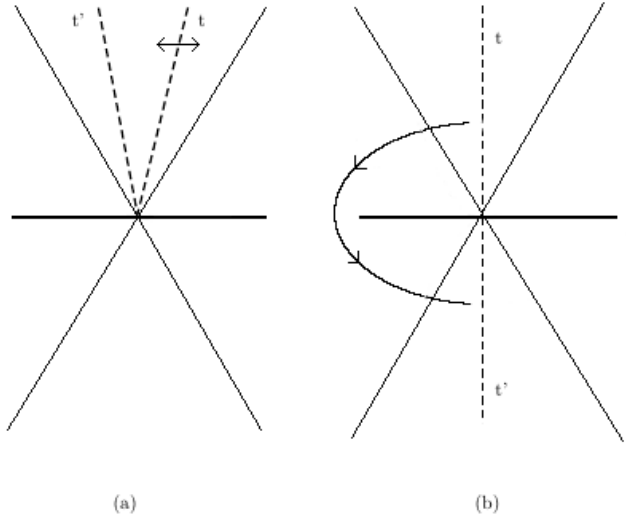}
\end{figure}

The standard definition of the time-reversal consists of 1) change of the sign of the time variable 2) interchange of the initial and final states. Then the time-reversal operation $ \hat{T} $ transforms a certain state of a particle to another state. The wave-function of a particle $ \psi_{p} $ with a certain 4-momentum $p$ cannot be the eigenfunction of $ \hat{T} $ with a certain eigenvalue unlike the case of space inversion operation $ \hat{P} $. The wave function of a particle is the eigenfunction of $ \hat{P} $ with the eigenvalue $ P $ (space parity)
\begin{eqnarray}
\hat{P}\psi_{p}=P\psi_{p}
,
\end{eqnarray}
where $ P=\pm 1 $. In the case of additionally extended Lorentz group (Fig. 1 (b)) the second condition in the definition of the time-reversal operation $\hat{T}$ is not necessary since the change of the sign of the time axis is allowed. The requirement to exchange the initial and final states was introduced by E. Wigner in \cite{13a} to restore the correct dependence of the particle's wave function on the time in the process of the time-reversal. In the case of the alternative interpretation, the same is achieved by the change of the direction of the time axis. Then the wave function of a particle can be made an eigenfunction of some operator $\hat{A}$ (see below) with eigenvalue $A$
\begin{eqnarray}
\hat{A}\psi_{A}=A \psi^{c}_{A}
\end{eqnarray}
where $ \psi^{c}_{A} $ is the charge-conjugated wave-function, $ A =\pm 1 $ corresponds to the direction of the time axis, "time arrow" and becomes the characteristics of every particle. This operator can be called the "time arrow operator". Eq. (2) is the main consequence of the alternative definition of the time-reversal in the relativistic theory.

Here we have to comment on the use of the term "time arrow". Commonly it is employed in a somewhat different sense. Introducing the alternative definition of the time-reversal, we have to distinguish between "mechanical" time which enters in all equations of motion, and "thermodynamical"
time which governs thermodynamical processes and decay processes of complex systems. The thermodynamical equations (like any conductivity equation) are not time-invariant and the decay processes  for complex systems are also time irreversible (see the recent discussion of the thermodynamical "time arrow" in \cite{12a,12b,12c,12d,12e}). The "thermodynamical" time, in principle, does not necessarily coincide with "mechanical" time and is not necessary for describing decay processes. It can be replaced by some decay characteristics. The most obvious choice is entropy, see \cite{12a, 12b, 12c, 12e}. As far as we know the term "time arrow" was used to fix the direction of "thermodynamical time", in particular, the direction of decay processes.

Unlike "thermodynamical", the "mechanical" time cannot be excluded from the theory (quantum or classical) since relativity loses its sense without time. In this communication, we consider only "mechanical" time, and our definition of the "time arrow" quantum number in Eq. (2) is pure "mechanical". Connection between the "mechanical" and "thermodynamical" time, though discussed in \cite{12a, 12b, 12c, 12e} requires further investigation.

Now let us consider how the time-reversal operator $\hat{T}_{A}$ corresponding to additionally extended Lorentz transformation acts on the field operators (for simplicity we consider the scalar field with particle's spin zero). The situation with spin $1$, $1/2$ particles differs mainly in the necessity to consider the behavior of polarization with time reversal. In this communication, we will be not interested in the polarization description.

Below we follow the considerations given in \cite{10} for the ordinary $\hat{T}$ operation. The field operator for the scalar field in the second quantization representation looks like
\begin{eqnarray}
\hat{\psi}(\textbf{r},t)=\sum\limits_{\textbf{p}}\frac{1}{\sqrt{2\epsilon}}\left[\hat{a}_{\textbf{p}}e^{-\mathrm{i}\epsilon t+\mathrm{i}\textbf{p}\textbf{r}} +
\hat{b}_{\textbf{p}}^{+}e^{\mathrm{i}\epsilon t-\mathrm{i}\textbf{p}\textbf{r}}
\right]
.
\end{eqnarray}
Here $\hat{a}_{\textbf{p}}$ is the particle annihilation operator, $\hat{b}_{\textbf{p}}^{+}$ is the anti-particle creation operator. Summation is extended over particle's momentum values $\textbf{p}$, $\epsilon$ is the particle's energy, $\textbf{r}$ and $t$ denote space an time coordinates, respectively. Expression Eq. (3) should be invariant under standard extended Lorentz transformation (including space, time inversion) with interchange of particles and antiparticles. For the creation and annihilation operators, this means
\begin{eqnarray}
\hat{a}_{\textbf{p}}\rightarrow \hat{b}_{\textbf{p}}^{+}
\\\
\nonumber
\hat{b}_{\textbf{p}}\rightarrow \hat{a}_{\textbf{p}}^{+}
.
\end{eqnarray}
On the other hand expression Eq. (3) should be 
invariant with respect to the CPT transformation. For the creation and annihilation operators in Eq. (3) P-transformation is
\begin{eqnarray}
\hat{P}: \hat{a}_{\textbf{p}}\rightarrow \pm \hat{a}_{-\textbf{p}}
\\\nonumber
 \hat{b}_{\textbf{p}}\rightarrow \pm \hat{b}_{-\textbf{p}}
 ,
\end{eqnarray}
and C-transformation is \cite{10}
\begin{eqnarray}
\hat{C}: \hat{a}_{\textbf{p}}\rightarrow \hat{b}_{\textbf{p}}
\\\nonumber
 \hat{b}_{\textbf{p}}\rightarrow \hat{a}_{\textbf{p}}
 .
\end{eqnarray}
Then from the condition that  CPT transformation coincides with extended Lorentz transformation for Eq. (3), for standard $\hat{T}$ transformation it follows
\begin{eqnarray}
\hat{T}: \hat{a}_{\textbf{p}}\rightarrow \pm\hat{a}_{-\textbf{p}}^{+}
\\\nonumber
 \hat{b}_{\textbf{p}}\rightarrow \pm\hat{b}_{-\textbf{p}}^{+}
 .
\end{eqnarray}
The transformation Eq. (7) has an evident physical sense \cite{10}: with the time reflection the momentum of the particle changes its sign and the beginning of the process is interchanged with its end.

Now we turn to the additionally extended Lorentz transformation when the time axis changes its sign. With this transformation in Eq. (3) the signs change twice: once due to the change the sign of variables $\textbf{p}$, $t$ and then due to the change of the time axis direction. As a result, the creation and annihilation operators in Eq. (3) remain intact: instead of Eq. (4) we formally write:
\begin{eqnarray}
\hat{a}_{\textbf{p}}\rightarrow \hat{a}_{\textbf{p}},
\\\nonumber
\hat{b}_{\textbf{p}}\rightarrow \hat{b}_{\textbf{p}}
\end{eqnarray}  
Now we have to choose the transformation $\hat{T}_{A}$ in the form, necessary to obtain Eq. (8) in combination with Eqs. (5), (6). This choice evidently is:
\begin{eqnarray}
\hat{T}_{A}:\; \hat{a}_{\textbf{p}}\rightarrow \pm\hat{b}_{-\textbf{p}},
\\\nonumber
\hat{b}_{\textbf{p}}\rightarrow \pm\hat{a}_{-\textbf{p}}.
\end{eqnarray}
The physical sense of the transformation $ \hat{T}_{A} $ is clear from Eq. (9): the particle's momentum changes its sign and the particles and antiparticles interchange with each other.
Now it becomes evident how the operator of the "time arrow"  $ \hat{A} $ in Eq. (2) should look like
\begin{eqnarray}
\hat{A}=\hat{T}_{A}=\hat{C}\hat{P}
\end{eqnarray}
 The signs $\pm$ in Eq. (9) correspond to the two values of the quantum number $ A $ in Eq. (2) and denote the two possible directions of the time axis ("time arrow") in our alternative interpretation of the time-reversal. 
 
It is easy to check that the operator $\hat{A}$ commutes with the Hamiltonian \cite{10}
\begin{eqnarray}
\hat{H}=\sum_{\textbf{p}}\epsilon [\hat{a}_{\textbf{p}}^{+} \hat{a}_{\textbf{p}} + \hat{b}_{\textbf{p}}^{+} \hat{b}_{\textbf{p}} ]
\end{eqnarray}
\begin{eqnarray}
[\hat{H}\hat{A}]_{-}=0
,
\end{eqnarray}
i. e. the "time arrow" $A$ can be really considered as the additional quantum number which characterizes the properties of particles. This can be done simply by changing the notation of summation variable in Eq. (10) $\textbf{p}=-\textbf{p}$.

Tracing the cosmological history for particles with both time arrow values ($A =\pm 1$) it is convenient to choose the Big Bang (BB) time as zero point for both directions of the time arrow. 

It is also natural to suppose that at the creation of the universe, particles with both time arrow values did arrive, not necessarily in equal amounts and not necessarily in amounts coinciding with amounts of participles and antiparticles. We can also think that the particles with the time arrow opposite to the time arrow in our universe formed another universe similar to our one. Though, according to Eq. (2) the particles in the other universe differ from the particles in our universe only by the time arrow, we will call this universe as anti-universe. 

The events, occurring in the anti-universe after BB (according to the anti-universe time arrow),
according to our time arrow has happened before the BB. In general, the future in the anti-universe (according to its time arrow) corresponds to the past, according to the time arrow of our universe, and vice versa: our future is the past with respect to the time arrow of the anti-universe. 


To understand whether the interactions between two universes are possible, we consider first the electromagnetic interactions. In principle, the whole construction of quantum electrodynamics (QED) remains unchanged. 

The time variable $t$ changes from $t=-\infty$ to $t=+\infty$ as in the standard QED. All the derivations and formulas remain the same including the renormalization scheme. Only the new quantum number $A = \pm 1$ is added. We consider first the process of emission (absorption) of the photon with arbitrary time arrow by an atomic electron also with arbitrary time arrow. For this purpose we will use the standard Feynman techniques with additional indices corresponding to the time arrows of the particles. The process of the photon emission  by an atomic electron will be described by the Feynman graph Fig. 2. The introduction of the time arrow quantum number $ A $ and the time arrow variable $\tau$ is similar to the introduction of particle spin and spin variables.

\begin{figure}[hbtp]
\caption{The Feynman graph corresponding to the process of photon emission (absorption) in the theory with an alternative description of the time-reversal. The solid lines as in the standard QED denote the electrons (positrons). The wavy line denotes the photon. The indices $A$ correspond to the time arrow quantum numbers. The indices for all other quantum numbers are omitted. The indices $x$, $ \tau $ at the vertex refer to the space-time variable $ x=(t,\textbf{r}) $ and to the "time arrow variable" $ \tau $. The integration over $ x $ and summation over $ \tau $ is assumed.}
\centering
\includegraphics[scale=0.5]{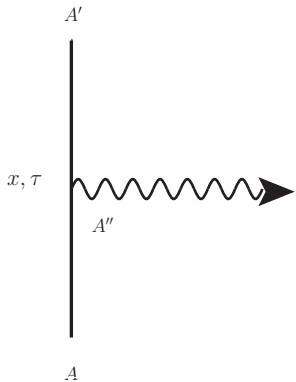}
\end{figure}

We present the wave function of any particle in the form 
\begin{eqnarray}
\label{wf}
\psi_{A}(x,\tau)=\Psi(x)\eta_{A}(\tau)
,
\end{eqnarray}
where $ \eta_{A}(\tau) $ is the component of spinor $ \eta_{A} $, the time arrow variable takes only two values $ \tau= \pm 1 $. It is convenient to choose 
\begin{eqnarray}
\label{kronecker}
\eta_{A}(\tau)=\delta_{A\tau} 
.
\end{eqnarray}
All other quantum numbers are omitted in Eq.(\ref{wf}). Then using the standard Feynman technique and retaining only the dependence on the time arrow variables, we can present the transition probability corresponding to the process Fig. 2 in the form 
\begin{eqnarray}
\label{waa}
W_{AA'} \sim \left|\sum_{\tau}\eta_{A'}(\tau)\eta_{A''}(\tau)\eta_{A}(\tau) \right|^2
.
\end{eqnarray}
Using the equality (\ref{kronecker}) we find
\begin{eqnarray}
W_{AA'} \sim \left|\sum_{\tau}\eta_{A'}(\tau)\eta_{A''}(\tau)\eta_{A}(\tau) \right|^2 = \delta_{A'A''}\delta_{A''A}
.
\end{eqnarray}
This means that the photon emitted in the anti-universe can not be absorbed (detected) in our universe and vise versa. Moreover, it is easy to show that any electromagnetic interaction between the particles of anti-universe  and the particles of the ordinary universe is absent. 

Here we have to mention that the various cosmological scenarios with extra universes and the backward time direction were already constructed by several authors for explaining the dark matter. The simultaneous creation of the universe and anti-universe via Big Bang was suggested in \cite{10a,10b}. In \cite{Alvarez:2018wst} a model was suggested where the dark matter was provided by the fields evolving back in time and acting in the regions with high metric curvature where the causality loses its sense. On the contrary, in \cite{Donoghue2020} it was argued that the causality is always present in the quantum field theory and defines the direction of time arrow. However, in all these works the time arrow operator Eq. (10) and its eigenfunctions and eigenvalues Eq. (2) were never introduced.

The anti-universe can be compared with the mirror universe, the existence of which was suggested in \cite{x1,x2}. This universe should contain right (R)-currents instead of the left (L)-currents in the existing Standard Model and was introduced to restore the balance between the right and left, i.e. P-invariance in nature. In this sense, we can say that the introduction of anti-universe could restore the CP (combined) invariance, suggested by Landau \cite{22a}. Both restorations, however, fail due to the inequality of the total amount of dark and ordinary matter. 

In \cite{x2} it was proved that the existence of electromagnetic interactions between L- and R- particles would contradict the experimental data of the decay of $\pi^{0}$ mesons. Therefore in \cite{x2} it was postulated that the electromagnetic (and as well strong) interactions between L- and R- particles should be prohibited. The gravitational interaction between L- and R- universes still remains. This makes the mirror universe one of the candidates to the role of the dark matter \cite{16}.

The electromagnetic interaction between the particles of the anti-universe and the ordinary universe is strictly prohibited (see Eqs. (15), (16)). Due to the similar structure of QED and QCD (quantum chromodynamics) theories, this should concern also strong interactions. Very small CP-violating weak interactions between two universes will be discussed at the end of this paper.

However, for the gravitational interactions between both universes, the situation is different. The interaction potentials between the two gravitating bodies following from the classical General Relativity (GR) depend only on the space, but not on the time intervals between these bodies \cite{x3}. Two gravitating bodies from two different universes are always divided by the time interval but may be located at the same space point, i.e. may interact with each other. In quantum gravity, the gravitons (quants of gravitation field) are massless tensor particles. Such particles may interact simultaneously with two different fields, one with positive time arrow and another with negative time arrow. Hence the arguments given above for the impossibility of electromagnetic (strong) interaction between the particles from different universes are not valid anymore.


Finally, we can try now to answer the question that was mentioned at the 
beginning of our paper: what happens with the particles after they undergo an interaction that reverses the time direction. We will discuss this question on the basis of an alternative interpretation of the time-reversal suggested in this communication. It is convenient to start with the discussion of the P, T-odd interaction of an atomic electron with the nucleus in a heavy atom (molecule). The Hamiltonian of this interaction looks like \cite{3,5,18}
\begin{eqnarray}
H_{P,T}=\mathrm{i}G_{F}g_{P,T}\gamma_{0}\gamma_{5}\rho(\textbf{r})
,
\end{eqnarray}
where $G_{F}$ is the Fermi constant, $g_{P,T}$ is a dimensionless constant depending on the particular model of the P, T-odd interaction \cite{8,9}, $\rho(\textbf{r}) $ is the distribution of the nuclear density, $\textbf{r}$ is the electron coordinate in an atom (molecule), $\gamma_{0} $ and $\gamma_{5} $ are the Dirac matrices. The factor $g_{P,T}$ defines the smallness of P, T-odd interaction, i.e. the level of CP violation. The CP violation within standard model (SM) originates from the phase factor in the Cabibbo-Kobayashi-Maskawa (CKM) matrix when all three generations of quarks are taken into account. With three generations this phase can not be excluded by any quark rotations. To keep this phase when calculating the amplitude of electron-nucleus P, T-odd interaction  in atoms it is necessary to consider 4-loop quark vertex on the electron line when using the two-photon exchange model \cite{8} or 3-loop vertex when using the Higgs boson exchange model \cite{9} for the scalar-pseudoscalar P, T-odd interaction between the electron and the nucleus in the atom (molecule).

It can be proved that the hermitian P, T-odd interaction Hamiltonian can be only of the scalar-pseudoscalar or tensor-pseudotensor type \cite{18}. The operator Eq. (17) is of the scalar-pseudoscalar type where the electron current is pseudoscalar ($\gamma_5$ matrix is acting on the electron variables). This contribution is dominant compared to the nuclear pseudoscalar contribution.
The tensor-pseudotensor interaction corresponds to the interaction of an electron (free or atomic one) EDM with an external electric field. In an external electric field, the interaction Eq. (17) produces the linear Stark shift of atomic levels the observation of which is the goal of modern experiments.

In this paper we are interested in another problem: whether it is possible to observe the P, T-odd effects connected directly with the interaction Eq. (17), in the absence of an external field. 

Since the operator $H_{P,T}$ changes the sign
of the time, the matrix element $\langle A (+1) | H_{P,T}|A(-1)\rangle$ should correspond to the transition between "contemporary" universes ($A=+1$) and ($A=-1$), i.e. between the world and anti-world at the same stage of development. Formally, using the algebraic representation Eq. (14) for the time arrow parts of wave functions $ \eta_{A} $ we can replace the factor $ i $  in Eq. (17), responsible for the time noninvariance, by the operator $i\equiv P(A= + 1 \rightleftarrows A=-1)$. The probability of such transition (transition rate) should look like
\begin{eqnarray}
W_{P,T}=\frac{2\pi}{m}|\langle A (+1) | H_{P,T}|A(-1)\rangle|^2
.
\end{eqnarray}
Eq. (18) is written for the transition of electron,
therefore $m$ in Eq. (18) is the electron mass. The factor $1/m$ in Eq. (18) is introduced for the proper normalization. The probability Eq. (18) corresponds to the process of spontaneous decay (ionization) of the ground state of an atom. Most probably the inner electron will undergo the transition, i.e. disappear (it transits to the anti-universe from our universe), since the nuclear density $\rho(\textbf{r})$ scales like $Z^3$ ($Z$ is the nuclear charge). The electron should disappear for the observer in our universe since as the result of transition it will become a particle in the anti-universe and will not interact anymore with the particles in our universe.

In principle, the same effect (disappearance of the electron due to the transition to another universe, i.e the charge exchange between two universes) may occur with the free electron; in an external electric field due to the tensor-pseudotensor interactions between the eEDM and electric field. However the P, T-odd interactions of bound electrons in heavy atoms (molecules) are strongly enhanced and the experiments with free electrons in electric fields are more difficult.

For the constant $g_{P,T}$ in Eq. (17) we can use an estimate for model interaction $H_{P,T}$ as an exchange by Higgs boson between the electron and the nucleus in an atom \cite{9}. This estimate reads: $g_{P,T}=10^{-9}$. Then in relativistic units ($\hbar=c=1$)
\begin{eqnarray}
\langle A=+1 |H_{P,T}|A=-1\rangle =10^{-9}G_{F}(m\alpha Z)^3
,
\end{eqnarray}
where $\alpha$  is the fine structure constant.
Employing the value of the Fermi constant $G_{F}=10^{-5}(m/m_p)^2$, where $m_p$ is the proton mass, setting $Z\sim 100$ and going over from r.u. to s$^{-1}$ we find
\begin{eqnarray}
W_{P,T}\approx 10^{-20}\,\mathrm{s}^{-1}
.
\end{eqnarray}

For the heavy atoms the process of the electron transition to the anti-universe should be accompanied by the X-ray emission. The observation of the small effects of spontaneous X-ray emission may be indistinct in the background of natural production of high energy photons by cosmic rays interactions with atoms etc. This situation is similar the proton decay in the hydrogen atom in huge water detector \cite{proton}. The expected rate of the proton decay is $W_{p}\approx 10^{-32}$ s$^{-1}$. The probability of the electron transfer to another universe in hydrogen ($Z=1$) is the same $W_{P,T}\approx 10^{-32}\,s^{-1}$. But in this case the effect can be detected only by the change of electric charge within the detector.     

The most important consequence of the proposed in this paper alternative interpretation of time-reversal is of cause the existence of anti-universe and the possibility to consider this universe as a source of dark matter. It is important also that in "principle" there is a possibility to confirm the existence of anti-universe in the laboratory experiments.

\bibliography{mybibfile}  

\end{document}